\documentclass[sigconf]{acmart}
\usepackage{amsmath}
\usepackage{algorithm}
\usepackage{algorithmic}
\usepackage{multirow}
\usepackage{xspace}
\usepackage{subfigure}
\usepackage{enumitem}

\newcommand{\method}{PAM\xspace}
\AtBeginDocument{%
  \providecommand\BibTeX{{%
    \normalfont B\kern-0.5em{\scshape i\kern-0.25em b}\kern-0.8em\TeX}}}

\copyrightyear{2025}
\acmYear{2025}
\setcopyright{acmlicensed}
\acmConference[KDD '25]{Proceedings of the 31st ACM SIGKDD Conference on Knowledge Discovery and Data Mining V.1}{August 3--7, 2025}{Toronto, ON, Canada}
\acmBooktitle{Proceedings of the 31st ACM SIGKDD Conference on Knowledge Discovery and Data Mining V.1 (KDD '25), August 3--7, 2025, Toronto, ON, Canada}
\acmDOI{10.1145/3690624.3709336}
\acmISBN{979-8-4007-1245-6/25/08}

\begin{document}

\title[Online Item Cold-Start Recommendation with Popularity-Aware Meta-Learning]{Online Item Cold-Start Recommendation with Popularity-Aware Meta-Learning}

\author{Yunze Luo}
\affiliation{
    \institution{School of CS, Peking University}
    \city{Beijing}
    \country{China}}
\email{lyztangent@pku.edu.cn}

\author{Yuezihan Jiang}
\affiliation{
    \institution{Kuaishou Technology}
    \city{Beijing}
    \country{China}}
\email{yuezihan.jiang@gmail.com }

\author{Yinjie Jiang}
\affiliation{
    \institution{Kuaishou Technology}
    \city{Beijing}
    \country{China}}
\email{jiangyinjie@kuaishou.com}

\author{Gaode Chen}
\affiliation{
    \institution{Kuaishou Technology}
    \city{Beijing}
    \country{China}}
\email{chengaode19@gmail.com}

\author{Jingchi Wang}
\affiliation{
    \institution{School of CS, Peking University}
    \city{Beijing}
    \country{China}}
\email{jamesw12@stu.pku.edu.cn}

\author{Kaigui Bian}
\authornote{Corresponding author.}
\affiliation{
    \institution{School of CS, Peking University}
    \city{Beijing}
    \country{China}}
\email{bkg@pku.edu.cn}

\author{Peiyi Li}
\affiliation{
    \institution{Kuaishou Technology}
    \city{Beijing}
    \country{China}}
\email{pyeeleezl@163.com}

\author{Qi Zhang}
\authornotemark[1]
\affiliation{
    \institution{Kuaishou Technology}
    \city{Beijing}
    \country{China}}
\email{zhangqi.cs.ucas@gmail.com}

\thanks{Yunze Luo, Jingchi Wang, and Kaigui Bian are affiliated with School of CS, AI Innovation Center, National Engineering Laboratory for Big Data Analysis and Applications, State Key Laboratory of Multimedia Information Processing, Peking University.}
\renewcommand{\shortauthors}{Yunze Luo et al.}

\begin{abstract}

With the rise of e-commerce and short videos, online recommender systems that can capture users' interests and update new items in real-time play an increasingly important role. In both online and offline recommendation systems, the cold-start problem caused by interaction sparsity has been impacting the effectiveness of recommendations for cold-start items. Many cold-start scheme based on fine-tuning or knowledge transferring shows excellent performance on offline recommendation. Yet, these schemes are infeasible for online recommendation on streaming data pipelines due to different training method, computational overhead and time constraints. Inspired by the above questions, we propose a model-agnostic recommendation algorithm called Popularity-Aware Meta-learning (PAM), to address the item cold-start problem under streaming data settings. PAM divides the incoming data into different meta-learning tasks by predefined item popularity thresholds. The model can distinguish and reweight behavior-related and content-related features in each task based on their different roles in different popularity levels, thus adapting to recommendations for cold-start samples. These task-fixing design significantly reduces additional computation and storage costs compared to offline methods. Furthermore, PAM also introduced data augmentation and an additional self-supervised loss specifically designed for low-popularity tasks, leveraging insights from high-popularity samples. This approach effectively mitigates the issue of inadequate supervision due to the scarcity of cold-start samples. Experimental results across multiple public datasets demonstrate the superiority of our approach over other baseline methods in addressing cold-start challenges in online streaming data scenarios.
\end{abstract}



\maketitle

\section{Introduction}
Platforms that depends on online recommendation service, including short video, e-commerce, and streaming media, play an indispensable role in people's lives. Online recommender systems need to capture interest shifts in the system through periodic updates while taking into account the historical information of users and items. Many previous works~\cite{devooght2015dynamic, SML, IncCTR} have demonstrated superior performance in the realm of online recommendation. These online systems are characterized by a large amount of data, strict time-consuming constraints, changes in global information over time~\cite{ASMG} and the requirement of one epoch streaming training. A common way to train the system streamingly is continuously optimizing the model to capture information shifts. The network parameters are applied for all users and items, and the parameters are updated using data arriving in the period to avoid being stale. 

Another inevitable recommendation scenario is the cold-start problem caused by sparse consumption data of new users or items. The fact that low-popularity items comprise a huge portion of the total items and their interaction data represents only a small fraction of the overall consumption data~\cite{KuaiRand, MovieLens} is also concerning. As a consequence of this long-tail distribution, the model excels in recommending popular items but struggles when it comes to recommending cold items. On the flip side, interactions with popular items provide a more accurate reflection of user interests and item features. Even recommendations for cold-start items heavily rely on the patterns learned by the model from these interactions.

Such cold-start problem has garnered considerable attention and has been addressed through various solutions in recommendation scenarios. The representation-based approach~\cite{MeLU,SSCDR,MWUF} considers items as tasks and generates personalized parameters for different items to enhance the effectiveness of cold-start item recommendation, such as meta-learning~\cite{MeLU} and knowledge transfer~\cite{MWUF,chen2023win, song2024mitigating, song2024multilora}. The side-information-based approach~\cite{Multi-modal,social,chen2024multi} focuses on using content-based side information and pay less attention to the item's popularity-related information, which also have different importance in the recommendation of cold and popular items. However, the methods above are still challenging to apply in online scenarios with the characteristic of streaming data.

The cold-start problem in online recommendation systems presents two main challenges. Firstly, when integrating online and cold-start recommendations, the Matthew effect causes the system to over-recommend popular items. This happens because popular items dominate the training data, leading to improved performance for these items and an increasing proportion of popular data over time. This effect exacerbates the cold-start problem in online systems. Secondly, the influx of numerous cold-start items meaks existing methods ineffective, as the time and computational demands make it impractical to generate personalized parameters for each cold-start item in online scenarios.

In this work, we have introduced a novel model-agnostic \textbf{P}opu-\\larity-\textbf{A}wared \textbf{M}eta-learning framework (\method) for solving the item cold-start problem under streaming data in online systems. Firstly, we propose a fixed task segmentation with predefined popularity thresholds in the gradient-based meta-learning~\cite{MAML}. We leverage meta-learning's ability to share meta-parameters between tasks and generate personalized parameters for different tasks. In this way, we optimise the performance of cold-start tasks without losing the interest information of streaming data, and the personalized parameters also avoid making the system over-fitting popular item data, solving the Matthew cold-start problem in online scenarios. In \method, tasks of varying popularity levels can share meta-knowledge and utilize popularity-related features and content-related features with different weights. Additionally, the fixed task segmentation enables efficient online inference without fine-tuning on each item by storing different tasks' parameters in advance, thus adapts to streaming training scenarios while addressing the computational overhead and time constraints.

Besides, we also design a cold-start task enhancer to further utilize various information from popular items in the system to optimize the cold-start task. The enhancer classifies the embeddings into behavior-based and content-based (because of their different roles in the recommendation of cold-start and popular items) and comprises two auxiliary tasks based on these two types of embeddings: 1) The data augmentation module constructs low-popularity samples using high-popularity interaction data, thereby directly increasing the number of cold-start samples; 2) the self-supervised module replay historical cold-start interactions leveraging well-trained item embeddings as supervision, thereby promoting the feature extraction capability for cold-start tasks.

Our main contributions can be summarized as follows:
\begin{itemize}[leftmargin=*]
\item We proposed a meta-learning algorithm that effectively tackles the item cold-start problem in online recommendation without requiring additional fine-tuning in online serving (Sec.~\ref{sec:serve}).
\item We introduced a unique cold-start enhancer that effectively mitigates the issue of scarcity feedback for cold-start items (Sec.~\ref{sec:enhancer}).
\item We conducted complete experiments on three benchmark datasets and a real online recommendation scenario. And our method surpasses previous baselines by a large margin.
\end{itemize}

\section{Related Works}
In this section, we discuss researches relevant to this work, including meta-learning, cold-start and online recommendation.

\subsection{Cold-Start Recommendation}

Solutions to the cold-start problem are usually categorized into side-information-based methods and fine-tuning-based methods~\cite{CIERec, cold-start2}. Cold-start recommendation methods based on side information are diverse~\cite{DropoutNet,CIERec,Multi-modal,HERec, cold-start1}. The more side-information is available to the system in online recommendation, the better the performance of such methods. Approaches based on fine-tuning are also suitable for cold-start scenarios, where features are extracted in advance through pre-training or meta-knowledge and can be quickly adapted to the task to achieve better results when faced with a new scenario with fewer samples~\cite{SSCDR}. In recent years, many hybrid methods that introduce side-information into few-shot learning have also emerged. As a method that utilizes a meta-learning framework to introduce content information to make recommendations for items of different popularity, \method is likewise a hybrid method that combines both. Meta-learning is also an efficient solution to adapt quickly to personalized parameters with a few samples, and we will describe the various methods in the subsequent subsections.

\subsection{Online Recommendation}
The goal of online recommendation is to assist users in discovering items they may be interested in real-time, and be able to distribute new items in the system promptly. Unlike sequential recommendations, which focus on the historical behavior of the system, online recommendations also have to take into account real-time interest shifts and be able to generate recommendation parameters in a low time-consuming process. As mentioned earlier, the large amount of online data and the streaming training method leads to many schemes that generate specialized parameters for sequential recommendation in offline systems to be infeasible. There are many types of common approaches, including CTR prediction based IncCTR~\cite{IncCTR} and DDP~\cite{DDP}, and FIRE~\cite{FIRE} that proposes a incremental recommendation algorithm from the perspective of graph signal processing. Various types of meta-learning algorithms introduced in the next subsection also shows potential in online recommendations. However, they still struggle with the long-tail distribution of online data and performance poorly on cold-start items.

\subsection{Meta-Learning}
Meta-learning~\cite{MAML}, also known as \textit{learning to learn}, is an emerging few-shot learning method that seeks to quickly adapt to the corresponding task with a few-shot data to achieve better recommendation results. The meta-learning training process divide the training set into support and query set, the former are used to generate unique parameters and the latter are used to calculate loss for training. Since MeLU~\cite{MeLU}, meta-learning has also demonstrated its applicability in the field of recommender systems due to the high overlap between few-shot learning and cold-start recommendation scenarios~\cite{LWA, MeLU, MAMO, MetaHIN, MetaCS, Mecos, DML, Clusterseq, PNMTA, M2EU}. In recent work, ClusterSeq~\cite{Clusterseq} preserves secondary user preferences through a sequential recommendation system based on meta-learning clustering. Online recommendation is also another scenario where the training process of meta-learning is applicable because of its characteristic of making recommendation model updates in a temporal order~\cite{SML, FORM, ASMG, LSTTM, MeLON, S2Meta}. As an example, SML~\cite{SML} trains a meta-model to output new parameters based on the parameters of the past moments and the data of the current moment. FORM~\cite{FORM} adapts user-specific tasks and leverages the online meta leader to optimize online training performance. 

\section{Preliminary}
\subsection{Notations}
In this work, we consider item cold-start recommendation task in streaming data scenario. 
The data arrives by time period $\{D_0, D_1, \cdots\!,$ $D_t, \cdots\}$, each period of data $D_t$ consists of user-item interactions $(u,i)\in\mathcal{U}\times\mathcal{I}$ that happened during $D_t$. The recommender system is similarly updated by time, with $\Phi_t, \Theta_t, \Omega_t$ representing different parts recommender at time $t$.
The system maintains the embedding matrix parameters $\Phi$ and the network weight parameters $\Omega$. We also define the initialization of the network weights as $\Theta$. Matrices and vectors are denoted by symbols in bold font. For a particular interaction data $(u,i)$, $v_i\in \mathbb{N}$ represents the number of views of the item, \textit{i.e.}, the number of times the item has been clicked by the user prior to this moment in time. In particular, for some pre-given threshold $v_{cold}$, items with popularity below that threshold are called cold-start items, and others are called popular items.

\subsection{Problem Statement}
\begin{figure*}[t]
    \centering
    \includegraphics[width=.9\linewidth]{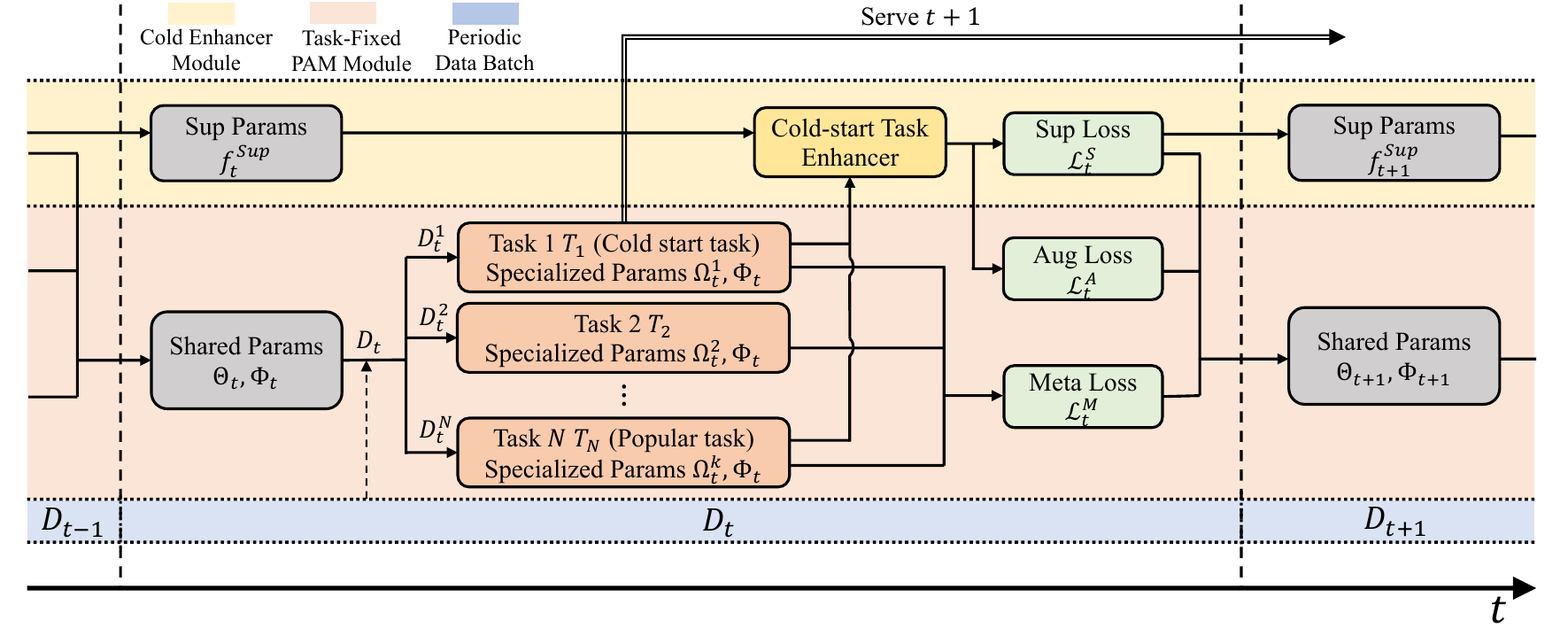}
    \caption{The overview of the proposed \method method. $\Phi_t$ denotes the task-shared embedding parameters, $\Theta_t$ denotes the task-shared network parameters initialization that will be fine-tuned to specialized parameters $\Omega_t^n$ using $D_t^n$ for each task.}
    \label{fig:overview}
\end{figure*}

Our goal is to consider the item cold-start problem in online recommender systems. In traditional online recommendation schemes, the system periodically retrains the system. To ensure that various types of historical data in the system are preserved and the real-time interest is captured, the generation of the current recommendation system $W_t$ needs to encapsulate the past information and search for better performance on the preceding data period:
\begin{equation}
    D_{t+1}\Leftarrow W_t\gets \mathcal{M}(W_{t-1}, \{D_i, i\leq t\})
\end{equation}
where $\mathcal{M}$ can be a gradient minimization or a meta generator, \textit{e.g.}, MeLON~\cite{MeLON} and SML~\cite{SML}. $\Leftarrow$ stands for serving. However, the above approaches suffer from the poor performance on cold-start items brought about by the scarcity of cold-start data. The cold-start online recommendation problem can be formulated as:
\begin{equation}
    D_{t+1}^{cold}\Leftarrow W_t\gets \mathcal{M}(W_{t-1}, \{D_i, i\leq t\})
\end{equation}
where $D^{cold}$ stands for the cold-start item part of data. Next, we present the \method method that addresses the problem of cold-start items in online recommender systems.

\section{Method}
In this section, we describe the proposed training method in detail, including the task-fixed meta-learning module, cold-start instruction module, and the base recommender model we adopt. The overview of the \method method is described in Fig.~\ref{fig:overview}.

\subsection{Base Recommender Model}
We would highlight that \method is a model-agnostic online training approach. We utilize a dual-tower structure~\cite{dual-tower} to ensure that it is uniform across all frameworks in this paper.

\subsubsection{Embedding layer} $(u,i)\to \boldsymbol{e}_u, \boldsymbol{e}_i$:
The purpose of the embedding layer is to characterize the different information about the user and the item, convert it into a vector, and output it to the hidden layer for information extraction. Specifically, the embedding layer consists of multiple feature embedding matrices of users and items. For user $u$, the system maintains multiple one-hot vectors for different features, which extract the user's information from the feature embedding matrix and then concatenate it into the embedding vector:
\begin{equation}
    \boldsymbol{e}_u = [\boldsymbol{e}_{u1}, \cdots, \boldsymbol{e}_{uP}] = [\boldsymbol{E}_{U1}\boldsymbol{c}_{u1}, \cdots, \boldsymbol{E}_{UP}\boldsymbol{c}_{uP}]
\end{equation}
where $P$ is the number of user features, $\boldsymbol{E}_{Up}\in\Phi_t$ represents the $p$-th feature embedding matrix and $\boldsymbol{c}_{up}$ stands for the one-hot vector for user feature $p$. Items have similar embedding outputs as users:
\begin{equation}
    \boldsymbol{e}_i = [\boldsymbol{e}_{i1}, \cdots, \boldsymbol{e}_{iQ}] = [\boldsymbol{E}_{I1}\boldsymbol{c}_{i1}, \cdots, \boldsymbol{E}_{IQ}\boldsymbol{c}_{iQ}]
\end{equation}

\subsubsection{Hidden layer}\label{sec:top} $\boldsymbol{e}_u, \boldsymbol{e}_i\to \boldsymbol{z}_u, \boldsymbol{z}_i$:
After the embedding vectors are generated, they are fed into the hidden layer and ultimately output the top representations of the user and item. In our dual-tower model, the user and the item have their own fully connected layer parameters, and the embedding vector will be mapped to the top representation by $L$ fully connected network layers $\boldsymbol{z}_u=f_{uL}\circ\cdots\circ f_{u1}(\boldsymbol{e}_u)$ and $\boldsymbol{z}_i=f_{iL}\circ\cdots\circ f_{i1}(\boldsymbol{e}_i)$, where the $l$-th layer can be represented as:
\begin{equation}
    f_{l}(\boldsymbol{e})=\sigma(\boldsymbol{W}_l\boldsymbol{e}+\boldsymbol{b}_l)
\end{equation}
where $\boldsymbol{W}_l, \boldsymbol{b}_l\in\Omega_t$ are the weight matrix and bias vector of the $l$-th hidden layer, and $\sigma$ represents a ReLU~\cite{ReLU} activation function.

\subsubsection{Output layer} $\boldsymbol{z}_u, \boldsymbol{z}_i\to\hat{y}_{ui}$:
The output layer calculates the predicted scores for the corresponding user-item pairs based on the top-level representations of the users and items. We utilize the InfoNCE~\cite{InfoNCE} loss form to calculate the prediction and loss value. The prediction of interaction $\hat{y_{ui}}$ is calculated by sampling other interactions in the batch as negative samples:
\begin{equation}
     \hat{y_{ui}} = \frac{\exp({\boldsymbol{z}_u \cdot \boldsymbol{z}_i/\tau})}{\sum_{u' \in D_t} \exp({\boldsymbol{z}_{u'} \cdot \boldsymbol{z}_i/\tau})}
\end{equation}
where $\tau$ is the temperature hyperparameter in InfoNCE loss to control the model's discrimination for negative samples. The log loss form is adopted to calculate the gradients:
\begin{equation}
     \mathcal{L}_{T} = - y_{ui}\cdot\log \hat{y_{ui}} - (1 - y_{ui}) \cdot\log (1 - \hat{y_{ui}})
     \label{equ:loss}
\end{equation}
where $\hat{y_{ui}}$ and $y_{ui}$ refer to the prediction and label values of $(u,i)$.

\subsection{Popularity-Aware Meta-Learning}
As the cold-start item samples constitute only a small fraction of the online traffic, their influence on model updating is limited. Consequently, the recommender system may exhibit satisfactory performance for popular items but may perform poorly for cold-start items. However, directly excluding samples from popular items could result in reduced performance for all items. This is because samples from popular items also contain user interest-related information, and the exclusion leads to an information loss. 

Hence, it naturally occurred to us to partition the tasks according to popularity and employ gradient-based meta-learning for recommendation. This popularity-aware segmentation enables us to strike a balance by avoiding excessive emphasis on highly popular items while still leveraging the valuable information from interactions with high-popularity items. Meanwhile, the segmentation is fixed during online training and it offers benefits from two perspectives. Firstly, it mitigates the overhead associated with traditional meta-learning patterns, which treat every data instance as a separate task, thus requiring task-specific parameter updating and storage. These methods incur unacceptable costs for online application. Secondly, as mentioned in~\cite{ANIL}, the meta-learner extracts versatile features for all tasks, and after fine-tuning on specific task, the model can reuse these features with different weights for fast adaption. In \method, items with the same popularity are recommended with identical features, while tasks with different popularity share lower-level embedding features. In other words, cold-start items can leverage more popularity-independent features, \textit{i.e.}, content features, while popular items can rely more on popularity-related features, \textit{i.e.}, historical feedback features.
\subsubsection{Fixed Task Segmentation}
Specifically, for a designated number of tasks $N$ and an interaction $(u, i)$, the number of tasks to which this interaction belongs is calculated by a piece-wise constant function:
\begin{equation}
    F(v_i): v_i\in\mathbb{N}\to T_n\in\{T_1,\cdots,T_N\}
\end{equation}
where $T_n$ represents the $n$-th task, and $v_i$ is a metric reflects popularity, such as the number of click or sales volume. The function $F$ is defined by a set of pre-determined thresholds. 

\subsubsection{Local Updates}
The parameters in \method follow a bi-level optimization scheme, consisting of local updates and global updates. 
Before the training process starts, we initialize the global recommender parameters $\boldsymbol{E}_{Up},\boldsymbol{E}_{Iq}\in\Phi_t$, $\boldsymbol{W}_l, \boldsymbol{b}_l\in\Theta_t$ . After the arrival of a data batch $D_t$, the batch will be divided into multiple parts $\{D_t^1, \cdots, D_t^N\}$ according to the popularity of the item by function $F$. Each portion of data $D_t^n$ is further divided into ${D_t^n}^S$ and ${D_t^n}^Q$, standing for support and query set.
For the $n$-th task, the personalized recommender parameters $\Omega_t^n$ can be obtained by local updates:
\begin{equation}
    \Omega_t^n\gets\Theta_t - \alpha\nabla_{\Theta_t}\mathcal{L}_{T_n}(\Theta_t,\Phi_t^n|{D_t^n}^S)
    \label{equ:l_update}
\end{equation}
where $\alpha$ represents the inner loop learning rate, and $\mathcal{L}_T(\Theta|D)$ represents the loss function of task $T$ on data set $D$ with an initialization of $\Theta$ (see Eq.~(\ref{equ:loss})). The recommender $\{\Phi_t, \Omega_t^n\}$ will be used to serve the items in task $n$ at the following time moment $t+1$. Note that the interaction data between the support set and the query set have no overlap, hence it is meaningless to update the embedding parameters of users and items in the local update, and the updated parameters contain only weight parameters of the networks in Eq.~(\ref{equ:l_update}). We utilize the LSLR~\cite{LSLR} methods, maintaining an adaptive learning rate for each network weight to alleviate overfitting and advance the performance.

\subsubsection{Global Updates}
The goal of the meta optimization is to minimize the sum of the losses of the query set ${D_t^n}^Q$ over its parameters $\Omega_t^n$ in each task $T_n$. Since all of the parameters are involved in the computation of the target function, the global update will be performed on all of the parameters: 
\begin{equation}
    \{\Phi_{t+1},\Theta_{t+1}\}\gets\{\Phi_t,\Theta_t\} - \beta\nabla_{\{\Phi_t,\Theta_t\}}\mathcal{L}^{M}_t
    \label{equ:g_update}
\end{equation}
where $\mathcal{L}^{M}_t$ is the main meta-learning loss, $\beta$ stands for the outer loop learning rate, and
\begin{equation}
    \mathcal{L}^{M}_t=\sum_{n=1}^N\lambda_n\mathcal{L}_{T_n}(\Omega_t^n,\Phi_t^n|{D_t^n}^Q)
    \label{equ:loss_m}
\end{equation}
where $\lambda$ represents the personalized weight of tasks. Because of the fixed division of the tasks, we can leverage the importance of the task and assign distinctive weights. For instance, the cold-start task, involving less popular items, can be assigned a higher weight.

By fixed tasks division as described above, we solve the problem of long-tailed distribution of training data by dividing the cold-started long-tailed part of the item distribution into separate tasks.

\begin{figure*}[t]
    \centering
    \includegraphics[width=\linewidth]{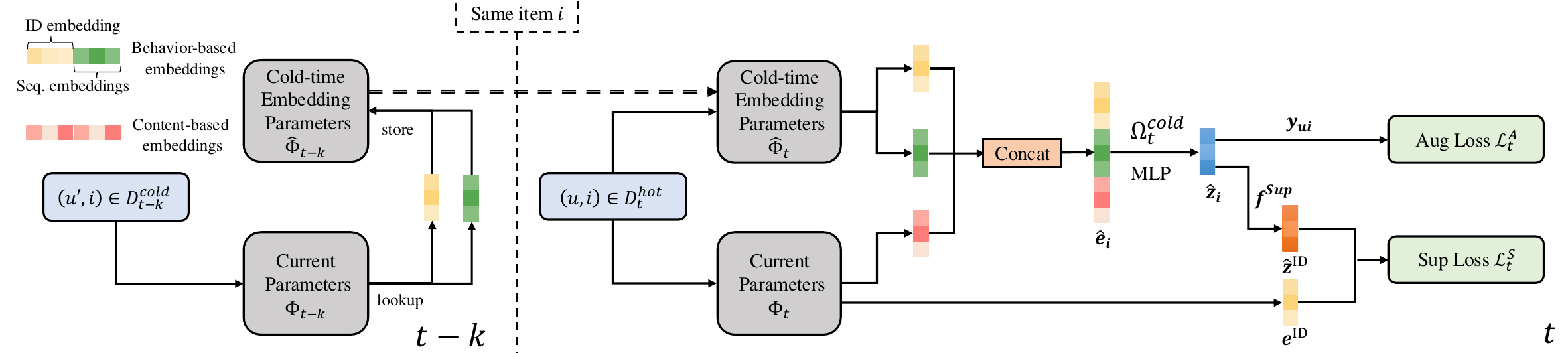}
    \caption{The structure of our proposed cold-start task enhancer.}
    \label{fig:instructor}
\end{figure*}

\subsection{Cold-start Task Enhancer}
\label{sec:enhancer}
Apart from task partitioning in meta-learning, incorporating additional supervision signals can prevent the neglect of cold-start tasks during online training. Thus, we devise a cold-start task enhancer that effectively transfers information from popular items and their associated interaction feedback. This module further improve the recommendations in the cold-start task.

\subsubsection{Cold-start Embedding Simulation} 
We categorize the various types of feature embedding of items into two types: behavior-based embeddings $\boldsymbol{e}_q^{beh}$ and content-based embeddings $\boldsymbol{e}_q^{con}$, because these types of embeddings play different roles in the recommendation of cold-start and popular items, as shown in Sec.~\ref{sec:breakdown}. 
We then further divide the behavior-based embeddings into ID embedding and sequential embeddings, this is because that ID embedding directly stores the information of the items and we utilize it for the cold-start embedding simulation, the details of which will be described later.
The system maintains a cold-start embedding parameter $\hat{\Phi}$, for each item $i$, whenever it appears in the data $D_{t-k}^{cold}$ as a cold-start item, the system stores its behavior-based embeddings (ID embedding and sequential embeddings) into the parameter $\hat{\Phi}_{t-k}$ in addition to updating its own embedding parameter $\Phi_{t-k}$. For each item $i$ that appears in the popular task $D_{t-k}^{hot}$, its behavior-based embeddings in the cold-start period have certainly been stored. Therefore, we can simulate a cold-started item at the current moment with the help of the stored embedding $\hat{\boldsymbol{e}}_{q}^{beh}\in\hat{\Phi}_{t-k}$ as well as other content-based feature embedding $\boldsymbol{e}_q^{con}\in\Phi_t$:
\begin{equation}
    \hat{\boldsymbol{e}}_i = [\hat{\boldsymbol{e}}_{i1}^{beh}, \cdots, \hat{\boldsymbol{e}}_{ik}^{beh}, \boldsymbol{e}_{ik+1}^{con}, \cdots, \boldsymbol{e}_{iQ}^{con}]
\end{equation}
The equivalent write-up for dividing behavior-based embeddings $\hat{\boldsymbol{e}}_{q}^{beh}$ into ID embedding $\hat{\boldsymbol{e}}_{q}^{ID}$ and sequential embeddings $\hat{\boldsymbol{e}}_{q}^{seq}$ is:
\begin{equation}
    \hat{\boldsymbol{e}}_i = [\hat{\boldsymbol{e}}_{i}^{ID}, \hat{\boldsymbol{e}}_{i2}^{seq}, \cdots, \hat{\boldsymbol{e}}_{ik}^{seq}, \boldsymbol{e}_{ik+1}^{con}, \cdots, \boldsymbol{e}_{iQ}^{con}]
    \label{equ:partemb}
\end{equation}
and $\hat{\boldsymbol{e}}_i$ fully simulates a cold-start item embedding at current time.

\subsubsection{Data Augmentation}
With relatively low traffic proportion, the cold-start task in meta-learning still has limited training opportunity. Data augmentation can significant increase the number of cold-start samples, so we propose a simulation-based data augmentation.
We treat the simulated cold-start embedding $\hat{\boldsymbol{e}}_i$ from the previous subsection as a new part of cold-start task data with the real popular interaction data labels $y_{ui}$, and train the meta-learning model to update the cold-start task parameters, which can partially augment the data to improve the effect.

We denote the simulated cold-start interaction data as $\hat{D}_t^{cold}$ and similarly divide them into support and query sets ${\hat{D}^{cold,S}_t}$, ${\hat{D}^{cold,Q}_t}$. Similar to Eq.~(\ref{equ:g_update}), the loss function form of data augmentation is computed from the data in the support set:
\begin{equation}
    \mathcal{L}^{A}_t=\mathcal{L}(\hat{\Omega}_t^{cold}|\hat{D}_t^{cold,Q})
    \label{equ:loss_a}
\end{equation}
where $\hat{\Omega}_t^{cold}$ is calculated by support sets ${\hat{D}^{cold,S}_t}$.

\subsubsection{Self-supervised Instructor}
Furthermore, we propose leveraging the well-learned embeddings of popular items and achieve instruction for the cold-start task by training a mapping network to fit the transformation process of the cold-start embedding to the embedding from popular phase. For the simulated cold-start item embedding $\hat{\boldsymbol{e}}_i$, we have the ID embedding of its popular state $\boldsymbol{e}^{\mathrm{ID}}\in\Phi_t$, which has more information due to multiple updates. To this end, we reinforce the ability of the other fully connected layers to extract information by replacing the last fully connected layer of the cold-start task network parameters $\boldsymbol{W}_L, \boldsymbol{b}_L$ with a unique mapping parameter $\boldsymbol{W}^{Sup}, \boldsymbol{b}^{Sup}$ that makes the output top embedding as similar as possible to its true ID embedding:
\begin{equation}
    \hat{\boldsymbol{z}}^{\mathrm{ID}}=f^{Sup}(\boldsymbol{e}_{L-1})=\boldsymbol{W}^{Sup}\boldsymbol{e}_{L-1}+\boldsymbol{b}^{Sup}
\end{equation}
where $\hat{\boldsymbol{z}}^{\mathrm{ID}}$ is the output embedding. We use MSE loss to evaluate the similarity between output top embedding and true ID embedding:
\begin{equation}
    \mathcal{L}^{S}_t = \frac{1}{N}\Vert\hat{\boldsymbol{z}}^{\mathrm{ID}}, \boldsymbol{e}^{\mathrm{ID}}\Vert_2^2
    \label{equ:loss_s}
\end{equation}
where $N$ stands for the dimension of embedding. 

With the instruction of the popular ID embedding, we can implement a self-supervised learning-like method without the use of labels to make the cold-start task network parameters have better performance in extracting information.

\begin{algorithm}[t]
\caption{The training and serving process of \method.}\label{alg:main}
\begin{algorithmic}
\REQUIRE Hyperparameters $\alpha$, $\beta$, $\lambda$, $\gamma$, $F$
\STATE Randomly initialize parameters $\Phi_0$, $\Theta_0$, $f^{Sup}_0$
\WHILE{incoming periodic data $D_t$}
\STATE $\{D_t^1,\cdots,D_t^N\}\gets F(D_t)$
\COMMENT{Divide data by popularity}
\FOR{$n\in\{1,\cdots,N\}$}
\STATE ${D_t^n}^S,{D_t^n}^Q\gets D_t^n$
\STATE $\Omega_t^n\gets\Theta_t - \alpha\nabla_{\Theta_t}\mathcal{L}_{T_n}(\Theta_t,\Phi_t^n|{D_t^n}^S)$
\ENDFOR
\STATE $\mathcal{L}^{M}_t\gets\sum_{n=1}^k\lambda_n\mathcal{L}_{T_n}(\Omega_t^n,\Phi_t^n|{D_t^n}^Q)$
\COMMENT{Meta-learning loss}
\IF{$t\geq\tau$}
\STATE Store cold-start parameter $\Omega_{t}^{cold}$ 
\STATE Serve $D_t^{cold}$ using $\Phi_{t-1}$, $\Omega_{t-1}^{cold}$ 
\COMMENT{Serving period}
\ENDIF
\STATE Store $\Phi_t$ of cold-start part $D_t^{cold}$ into $\hat{\Phi}_t$
\STATE Concatenate $\hat{\boldsymbol{e}}_i$ using $\hat{\Phi}_t,\Phi_t$ of hot part $D_t^{hot}$
\STATE Lookup $\boldsymbol{e}^{\mathrm{ID}}_i$ using $\Phi_t$ of hot part $D_t^{hot}$
\STATE Calculate $\hat{\boldsymbol{z}}^{\mathrm{ID}}$ using $\hat{\boldsymbol{e}}_i$ and network parameters $\Omega_t^{cold},f_t^{Sup}$
\STATE $\mathcal{L}^{S}_t\gets\frac{1}{N}\Vert\hat{\boldsymbol{z}}^{\mathrm{ID}}, \boldsymbol{e}^{\mathrm{ID}}\Vert_2^2$
\COMMENT{Self-supervised instructor loss}
\STATE Form a simulated cold task ${\hat{D}^{cold,S}_t},{\hat{D}^{cold,Q}_t}$ using $\hat{\boldsymbol{e}}_i,D_t^{hot}$
\STATE $\hat{\Omega}_t^n\gets\Theta_t - \nabla_{\Theta_t}\alpha_{\Theta_t}\mathcal{L}(\Theta_t,\hat{\Phi}_t^n|{\hat{D}_t^{cold,S}})$
\STATE $\mathcal{L}^{A}_t\gets\mathcal{L}(\hat{\Omega}_t^{cold}|\hat{D}_t^{cold,Q})$
\COMMENT{Data augmentation loss}
\STATE $\mathcal{L}^{T}_t\gets\gamma_M\mathcal{L}^{M}_t + \gamma_S\mathcal{L}^{S}_t + \gamma_A\mathcal{L}^{A}_t$
\COMMENT{Final total loss}
\STATE $\{\Phi_{t+1},\Theta_{t+1}, f^{Sup}_{t+1}\}\gets\{\Phi_t,\Theta_t,f^{Sup}_t\} - \beta\nabla_{\{\Phi_t,\Theta_t,f^{Sup}_t\}}\mathcal{L}^{T}_t$
\ENDWHILE
\end{algorithmic}
\end{algorithm}

\subsection{Training and Serving}
\subsubsection{Training Process}
With the introduction of the parts of the cold-start task enhancer, the update method of the recommender parameters in Eq.~(\ref{equ:g_update}) changes. We form a new total loss function by combining the three components of main meta-learning loss, self-supervised learning loss, and data augmentation loss:
\begin{equation}
    \mathcal{L}^{T}_t = \gamma_M\mathcal{L}^{M}_t + \gamma_S\mathcal{L}^{S}_t + \gamma_A\mathcal{L}^{A}_t
\end{equation}
where $\gamma$ refers to the different weights of losses, and the new parameter update method becomes:
\begin{equation}
    \{\Phi_{t+1},\Theta_{t+1}, f^{Sup}_{t+1}\}\gets\{\Phi_t,\Theta_t,f^{Sup}_t\} - \beta\nabla_{\{\Phi_t,\Theta_t,f^{Sup}_t\}}\mathcal{L}^{T}_t .
\end{equation}

\subsubsection{Online Serving}
\label{sec:serve}
During the training process, the model is fine-tuned on the cold-start data to generate parameters for the cold-start task $\Omega_{t}^{cold}$. We save that parameter and serve it for the recommendation of cold-start items at the next moment. In this case, \method is able to store the parameters prepared for the cold-start item in advance without real-time personalized fine-tuning, thus avoiding the time cost of online fine-tuning altogether. Compared to traditional meta-learning, we avoid the time-consuming serving of personalization on each item, and also reduce the storage space overhead of maintaining the respective parameters for each item.

The training and serving process of \method is described by Alg.~\ref{alg:main}.

\section{Experiments}
We conducted experiments~\footnote{Our code is available at \url{https://github.com/Sycamoretail/PAM}.} to answer the following RQs: 
\textbf{RQ1}: How does \method's performance on online cold-start recommendations improve over existing frameworks? 
\textbf{RQ2}: How much of a performance enhancement do the various components of \method provide? 
\textbf{RQ3}: Why \method perform better on multitasking? 
\textbf{RQ4}: How do the various hyperparameters of \method affect the performance results?
\textbf{RQ5}: How does \method perform in online A/B testing?

\begin{table}[t]
    \caption{Overview of the datasets.}
    \label{tab:dataset}
    \resizebox{\columnwidth}{!}{%
    \begin{tabular}{c|ccccc}
    \toprule
    \textbf{Dataset} & \textbf{\#Users} & \textbf{\#Items} & \textbf{\#Inter.}& \textbf{\#Tags}&\textbf{Density}\\ \midrule
    \textbf{MovieLens}& 43,181& 51,142& 6,840,091& 20&3.09\textperthousand\\
    \textbf{Yelp}& 1,987,929& 150,346& 6,990,280& 1,312&0.02\textperthousand\\
    \textbf{Book}& 351,487& 581,717& 6,402,728& 2,808&0.03\textperthousand\\\bottomrule
    \end{tabular}%
    }
\end{table}

\subsection{Experimental Settings}
\subsubsection{Datasets}
We selected three public datasets that containing timestamp to ensure that the characteristic of online streaming data is simulated, including \textbf{MovieLens}~\cite{MovieLens}, \textbf{Yelp} and \textbf{Book}~\cite{Book}. Details of datasets are shown in appendix~\ref{sec:dataset}.

For the label processing of the datasets, the user ratings lie between 0 and 5, while we consider the interaction data with ratings above 3 as positive samples and the rest as negative samples. The information of datasets is conducted in Table~\ref{tab:dataset}.

\subsubsection{Baselines}
We compare our proposed \method with the following baseline, where all scenarios leverage the previously mentioned two-tower model structure and the feature inputs remain the same.

- \textbf{Periodical Fine-tuning (PF)} or incremental updating, is a frequently used type of update in online systems, which performs an update on recommender system using the incoming data periodically to serve the next moment.

- \textbf{s$^2$Meta~\cite{S2Meta}} employs a novel meta-learning task division approach that considers item scenarios as tasks to achieve better recommendation results for new scenarios.

- \textbf{IncCTR~\cite{IncCTR}} is a practical incremental method that leverages knowledge distillation for training, use the current knowledge stored in model to guide the retraining.

- \textbf{SML~\cite{SML}} proposes a meta-model that takes the previous moment's recommender and current moment's data as inputs and generate the next moment's recommender for serving.

- \textbf{ASMG~\cite{ASMG}} better captures long-term interests through historical recommenders to generate models for serving through a meta-model generator that introduces GRU.

- \textbf{MeLON~\cite{MeLON}} generates more stable online recommendation models by maintaining dynamic learning rates for different parameters and distinguishing the importance of interaction data.

- \textbf{IMSR~\cite{IMSR}} allows the traditional adaptive interest capture model to better capture new interests of users periodically by introducing unique interest shifters..

\subsubsection{Evaluation Metrics}
Similar to ASMG~\cite{ASMG}, we divide each dataset equally into 31 periods, and each period is further partitioned into batches for training. To simulate the characteristics of online streaming data, we do not use a pre-training scheme, and all models will be consistently streamed starting from the data of the first period. We start testing from the 24th period, and the models obtained from each training period will be tested on cold-start data from the next period. The average of the metrics from all the testing phases will be reported. For specific metrics, we use the Recall@K and NDCG@K~\cite{NDCG} metrics to measure the model's effectiveness on cold-start items. Since ranking the item's scores for all users is time-consuming, for a cold-start item, we take the interacted user as 1 positive sample and all the non-interacted users in the current batch as the negative samples (not less than 900 due to the item's low popularity), and compute the above metrics.

\subsubsection{Hyperparameters}
For the definition of the tasks, to simulate an online cold-start scenario, we designated the items with the lowest 5\% of popularity in the interaction data as cold-start items~\cite{CIERec, MAMO}. The cold-start thresholds of popularity $v_{cold}$ on the three datasets are 50, 20, and 15, respectively. We divide the remaining data into 4 tasks to distinguish popular items of different natures.

During the training process, we use Adam~\cite{Adam} as the optimizer for gradient descent with $\alpha$ set to 0.001. The outer loop learning rate $\beta$ is set to 0.001. The batch size is 1024, the weights of tasks $\beta$ are set to 2 for the cold-start task and 0.5 for the other tasks, and the weights of the loss function $\gamma$ are set to 1, 3, and 2, respectively.

\subsection{Overall Performance (RQ1)}

\begin{table*}[t]
\centering
\caption{The reported top-K evaluation metric results on cold-start items. Bold represents the optimal result, underlined represents the optimal result in baseline, and Impr. \% represents the improvement compared to the optimal result in baseline.}
\label{tab:perf}
\resizebox{\textwidth}{!}{%
{\small
\begin{tabular}{cccccccccccccc}
\toprule
\multicolumn{1}{c|}{Dataset} & \multicolumn{1}{c|}{Metric} & PF & s$^2$Meta& IncCTR & SML & ASMG & MeLON &\multicolumn{1}{c|}{IMSR } & \method-M & \method-S & \method-A & \multicolumn{1}{c|}{\method-F} & Impr. \% \\ \midrule
\multirow{6}{*}{\textbf{MovieLens}} & \multicolumn{1}{|c|}{Recall@5} & 0.2581& 0.2344& 0.2685& 0.1965& 0.1945& \underline{0.3132} &\multicolumn{1}{c|}{0.2115} & 0.3846& 0.3968& 0.4114& \multicolumn{1}{c|}{\textbf{0.4157}} & +32.73\%\\
& \multicolumn{1}{|c|}{Recall@10} & 0.3268& 0.2750& 0.3383& 0.2628& 0.2575& \underline{0.3925} &\multicolumn{1}{c|}{0.2885} & 0.4733& 0.4856& 0.4980& \multicolumn{1}{c|}{\textbf{0.5036}} & +28.31\%\\
 & \multicolumn{1}{|c|}{Recall@20} & 0.4121& 0.3342& 0.4221& 0.3470& 0.3404& \underline{0.4862} &\multicolumn{1}{c|}{0.3846} & 0.5727& 0.5840& 0.5912& \multicolumn{1}{c|}{\textbf{0.5966}} & +22.71\%\\
\cmidrule{2-13}
 & \multicolumn{1}{|c|}{NDCG@5} & 0.2011& 0.1967& \underline{0.2100}& 0.1448& 0.1427& 0.2009 &\multicolumn{1}{c|}{0.1668} & 0.3040& 0.3162& 0.3266& \multicolumn{1}{c|}{\textbf{0.3314}} & +57.81\%\\
 & \multicolumn{1}{|c|}{NDCG@10} & 0.2233& 0.2060& 0.2326& 0.1661& 0.1630& \underline{0.2429} &\multicolumn{1}{c|}{0.1769} & 0.3327& 0.3450& 0.3547& \multicolumn{1}{c|}{\textbf{0.3598}} & +48.13\%\\
  & \multicolumn{1}{|c|}{NDCG@20} & 0.2448& 0.2180& \underline{0.2813}& 0.1873& 0.1839&  0.2704&\multicolumn{1}{c|}{0.1816} & 0.3578& 0.3698& 0.3782& \multicolumn{1}{c|}{\textbf{0.3833}} & +36.26\%\\
\midrule
\multirow{6}{*}{\textbf{Yelp}} & \multicolumn{1}{|c|}{Recall@5} & 0.1264& \underline{0.1375}& 0.1350& 0.0852& OOM& 0.1253 &\multicolumn{1}{c|}{0.0918} & 0.1969& 0.2184& 0.2203& \multicolumn{1}{c|}{\textbf{0.2262}} & +64.51\%\\
& \multicolumn{1}{|c|}{Recall@10} & 0.2106& 0.2206& \underline{0.2247}& 0.1513& OOM& 0.2090 &\multicolumn{1}{c|}{0.1836} & 0.2966& 0.3261& 0.3244& \multicolumn{1}{c|}{\textbf{0.3316}} & +47.57\%\\
 & \multicolumn{1}{|c|}{Recall@20} & 0.3324& 0.3390& \underline{0.3496}& 0.2574& OOM& 0.3327 &\multicolumn{1}{c|}{0.2755} & 0.4199& 0.4512& 0.4472& \multicolumn{1}{c|}{\textbf{0.4538}} & +29.81\%\\
\cmidrule{2-13}
 & \multicolumn{1}{|c|}{NDCG@5} & 0.0807& \underline{0.0876}& 0.0858& 0.0528& OOM& 0.0791 &\multicolumn{1}{c|}{0.0631} & 0.1313& 0.1466& 0.1467& \multicolumn{1}{c|}{\textbf{0.1525}} & +74.09\%\\
 & \multicolumn{1}{|c|}{NDCG@10} & 0.1077& 0.1129& \underline{0.1145}& 0.0740& OOM& 0.1062 &\multicolumn{1}{c|}{0.0911} & 0.1633& 0.1813& 0.1802& \multicolumn{1}{c|}{\textbf{0.1865}} & +62.88\%\\
  & \multicolumn{1}{|c|}{NDCG@20} & 0.1383& 0.1416& \underline{0.1459}& 0.1006& OOM& 0.1370 &\multicolumn{1}{c|}{0.1140} & 0.1944& 0.2129& 0.2112& \multicolumn{1}{c|}{\textbf{0.2173}} & +48.93\%\\
\midrule
\multirow{6}{*}{\textbf{Book}} & \multicolumn{1}{|c|}{Recall@5} & 0.2131& 0.2146& \underline{0.2173}& 0.2104& 0.2182&  0.2002&\multicolumn{1}{c|}{0.2117} & 0.2456& 0.2561& 0.2468& \multicolumn{1}{c|}{\textbf{0.2609}} & +20.06\%\\
& \multicolumn{1}{|c|}{Recall@10} & 0.2991& 0.3022& 0.3048& 0.2943& 0.3066&  \underline{0.3059}&\multicolumn{1}{c|}{0.2090} & 0.3460& 0.3583& 0.3478& \multicolumn{1}{c|}{\textbf{0.3616}} & +18.21\%\\
 & \multicolumn{1}{|c|}{Recall@20} & 0.4033& 0.4056& 0.4079& 0.3968& 0.4126&  \underline{0.4311}&\multicolumn{1}{c|}{0.3327} & 0.4614& 0.4753& 0.4667& \multicolumn{1}{c|}{\textbf{0.4806}} & +11.48\%\\
\cmidrule{2-13}
 & \multicolumn{1}{|c|}{NDCG@5} & 0.1475& 0.1495& \underline{0.1517}& 0.1461& 0.1519&  0.1334&\multicolumn{1}{c|}{0.1204} & 0.1714& 0.1789& 0.1715& \multicolumn{1}{c|}{\textbf{0.1824}} & +20.23\%\\
 & \multicolumn{1}{|c|}{NDCG@10} & 0.1752& 0.1777& \underline{0.1799}& 0.1731& 0.1804&  0.1646&\multicolumn{1}{c|}{0.1546} & 0.2038& 0.2119& 0.2040& \multicolumn{1}{c|}{\textbf{0.2149}} & +19.46\%\\
  & \multicolumn{1}{|c|}{NDCG@20} & 0.2015& 0.2038& 0.2059& 0.1990& 0.2071&  \underline{0.2170}&\multicolumn{1}{c|}{0.2129} & 0.2329& 0.2414& 0.2340& \multicolumn{1}{c|}{\textbf{0.2449}} & +12.86\%\\
\bottomrule
\end{tabular}}
}
\end{table*}

\subsubsection{Comparison of Cold Items}
Table ~\ref{tab:perf} demonstrates results of the top-K metrics on cold-start items for various methods. From the result of the experiment, we have the following observations.

Our proposed \method significantly outperforms other baselines on cold-start items. This demonstrates the effectiveness of \method's unique meta-learning scheme targeted at solving long-tailed distributions of item popularity in online streaming data scenarios. The \method performs better on smaller K-value metrics compared to other baseline methods, and the improvement is relatively small as the K-value increases. This suggests that the \method method can accurately recommend cold-start items to the users who favor it the most. In contrast, the baseline method ranks the positive-sample users relatively far down the list, as reflected in the considerable rise in the metrics as the K-value rises.

SML and ASMG methods focus on extracting information from historical recommender systems and balancing the weights of historical information with real-time interests to optimize recommendations at the next moment. However, because data distribution tends to be popular videos, focusing on historical information only aggravates the interest bias of the model. Besides, this model relies more on the efficacy of the initialized model, which results in poorer performance on cold-start items in the streaming no pre-training scheme. At the same time, ASMG has a high storage space overhead (leads to OOM on the Yelp dataset), which becomes another limitation for its application in online systems. IMSR is also an approach that balances periodic retraining of historical with current information, and thus has a similarly reliance on the efficiency of pre-training, and performs poorly in schemes without pre-training. In addition, the emphasis on historical information extraction also leads to a worse performance on cold-start recommendations.

IncCTR, as a knowledge distillation scheme that utilizes the model's self-supervised signals, also references historical information. Its low dependence on initialized parameters leads to a better fit to online scenarios and a slight improvement over PF, achieving state-of-the-art for cold-start items on some datasets. However, its way of generating additional labels increases the computational time consumption and partially reduces the model's efficiency.

s$^2$Meta is a scheme for task segmentation and meta-learning training of arriving data by categories of items. However, it only improves slightly relative to the PF baseline when no new item categories enter the system and performs slightly inferior on the content-information-rich MovieLens dataset, as the rest of the baseline can similarly achieve similar information extraction through content-based embedding.

MeLON performs superiorly on some recall metrics and worse on NDCG metrics. As a scheme that can adjust the learning rate in both directions on interactions and parameters to accommodate necessary samples, MeLON can better find the direction of gradient updates in less cold-start data and thus performs better in the MovieLens dataset. However, as the number of cold-start interactions increases in other datasets, its inability to find the commonality of cold-start items may bring inefficiency in updating.

\subsubsection{Comparison of Popular Items} 
Meanwhile, we compare the metrics of \method on popular items, the results are showed in Table~\ref{tab:popular}.

\begin{table}[t]
    \centering
    \caption{The reported top-K evaluation metric results on popular items on MovieLens dataset.}
    \begin{tabular}{ccccccc}
        \toprule
         & R@5 & R@10 & R@20 & N@5 & N@10 & N@20 \\ \midrule
        PF & 0.2949& 0.3532& 0.4285& 0.2474& 0.2661& 0.2850\\
        s$^2$Meta & 0.2962 & 0.3544 & 0.4287 & 0.2488 & 0.2674 & 0.2861 \\
        IncCTR& 0.2932& 0.3515& 0.4247& 0.2466& 0.2653&0.2838\\
        SML& 0.2729& 0.3335& 0.4049& 0.2263& 0.2459& 0.2643\\
        ASMG& 0.2659& 0.3249& 0.3999& 0.2210& 0.2399& 0.2588\\
        MeLON& 0.3205& 0.4261& 0.5139& 0.2442& 0.2689& 0.3173\\
        IMSR& 0.2914& 0.3581& 0.4388& 0.2317& 0.2566& 0.2821\\
        \method & 0.3340& 0.4031& 0.4840& 0.2771& 0.2994& 0.3198\\
        \bottomrule
    \end{tabular}
    \label{tab:popular}
\end{table}

It can be seen that the performance of the other baseline methods on popular items has improved compared to the cold-start items, reflecting the shift of the parameters toward the popular items due to more feedback on popular items. Meanwhile, \method does not perform as well on popular as on cold-start items, demonstrating that the preference of parameter settings for cold-start tasks successfully optimizes on cold-start items. Notably, \method still outperforms some baselines on popular items, demonstrating the superior performance of popular task parameters detached from the cold-start data for recommending items in the corresponding task.

\subsection{Ablation Study (RQ2)}
We compare \method with several variants: \textbf{\method-M} refers to the method without the cold-start task enhancer, \textbf{\method-S} and \textbf{\method-A} denote the method that adds only the self-supervision and data augmentation modules, respectively, and \textbf{\method-F} is the method that implements the complete cold-start task enhancer. 

We can see that both the optimization method for cold-start task information extraction with self-supervised signals and the data augmentation method for simulated cold-start data significantly enhance the \method on all datasets. This reflects the nature of \method's excellent task segmentation, which makes further optimization on a single task feasible.

For \method-S, it utilizes self-supervised signals using the ID embedding information of popular items to guide the cold-start task. Compared with \method-M, the improvement on the MovieLens dataset is relatively minor, while it has a significant optimization effect on the Yelp dataset. This may be because the high concentration of content-based features in the MovieLens dataset leads to the cold-start task itself being able to extract more information. In contrast, the decentralized content-based features in the Yelp dataset make the guidance approach more effective.

The \method-A approach leverages the idea of data augmentation. Simulating cold-start consumption data using historical behavior-based embedding of popular items delivers a significant optimization compared to \method-M across all datasets. It also shows that the sparseness of feedback on cold-start items is a significant and essential reason for the poor recommendation of cold-start items. Thus, the increase in data volume further enhances the task segmentation to improve the performance of cold-start items.

\method-F combines all the proposed optimization methods for the cold-start task and ultimately brings significant improvements in the cold-start item metrics compared to the baseline. It is worth noting that the improvement effect of \method-F is weakened compared with the combination of the respective improvement of \method-S and \method-A, which only incorporate a single loss. This is because both \method-S and \method-A optimize the cold-start task, but they use different methods and thus optimize in slightly different directions, thus weakening the effect of \method-F's enhancement.

\subsection{Breakdown Analysis (RQ3)}
\label{sec:breakdown}
\begin{figure}[t]
    \centering
    \includegraphics[width=.9\linewidth]{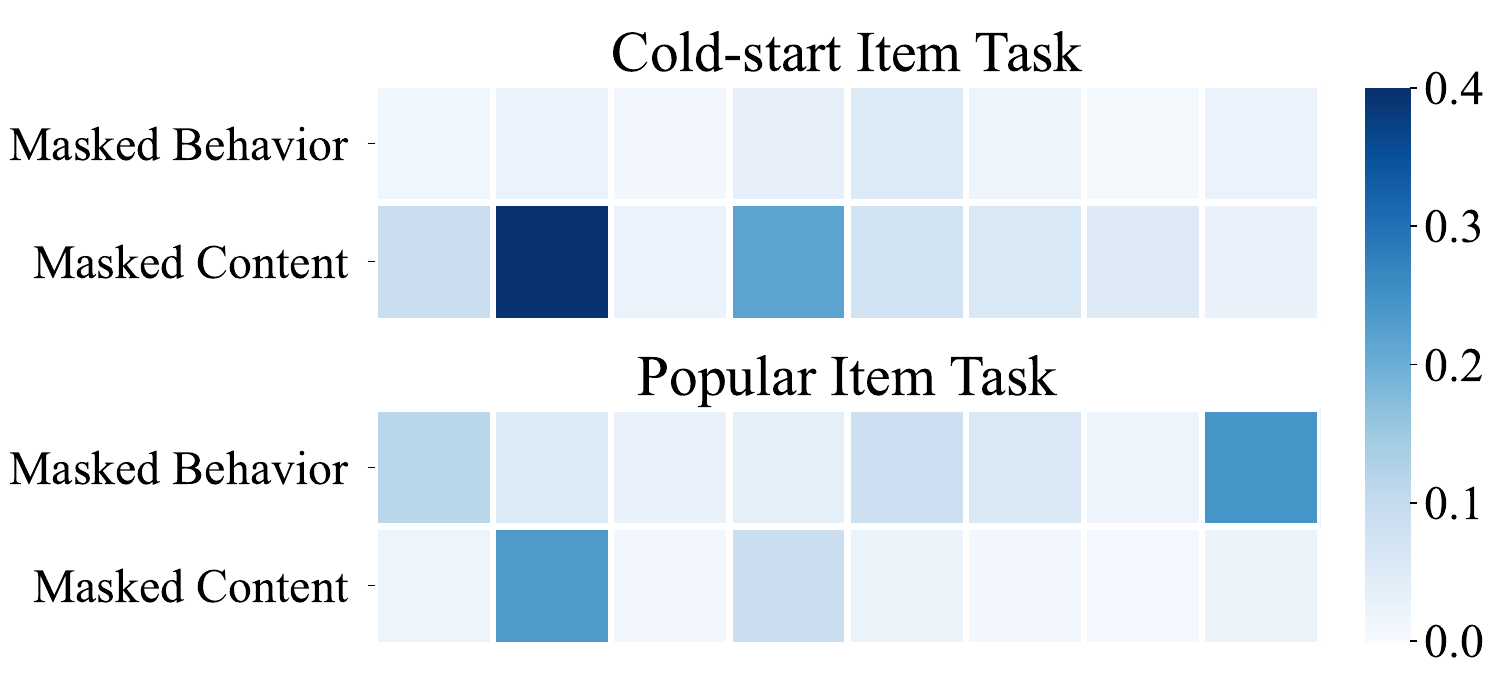}
    \caption{Squared errors of top representations of cold-start items and popular items before and after masking for different types of embedding inputs.}
    \label{fig:heat}
\end{figure}

In this section, we conduct a breakdown analysis of \method's ability to personalize parameters on tasks with different item popularity. To validate this, we summarize the following experiments. First, we saved the network parameters of cold-start and popular tasks generated by the corresponding task data. We sampled 200 batches and computed the top representations $\boldsymbol{z}_i$ in Sec~\ref{sec:top} of cold-start and popular items using the corresponding parameters. Afterward, we categorized the input embedding into two types like in Eq.~(\ref{equ:partemb}): behavior-based embedding $\boldsymbol{e}_q^{beh}$ and content-based embedding $\boldsymbol{e}_q^{con}$. We mask the behavior-based and content-based embedding of the cold-started and popular items to 0, respectively, and compute the top representations again as inputs using the corresponding network parameters. The masked top representations are further compared with unmasked top representations. Fig.~\ref{fig:heat} illustrates the squared error of the top representations of the cold-start and popular items before and after masking the two types of embeddings.

In the task of cold-start items, the differences in the top representations of the network outputs before and after masking the behavior-based embedding were minor. In contrast, some dimensions of the top representations changed considerably before and after masking the content-based embedding. In cold-start items, content-based embedding stores less information due to the few-shot interactions in cold-start items. Hence, the parameters generated by the cold-start task can effectively extract the content information, which gives more importance to content-based rather than behavior-based embedding and improves the recommendation effect of the cold-start task.

In contrast, the top representations of the popular item tasks changed considerably before and after both embedding masks, and masking the behavior-based embedding introduced more significant error. This suggests that the network parameters of the popular item task can effectively utilize the information in both embedding and focus more on behavior-based embedding (\textit{e.g.}, ID embedding).

We can also that masking the content-based embedding, both on the cold-start task and on the hot task, affects the exact same dimensions of the top-level representation. illustrating the similar extraction methods of content information of the cold-start and popular item tasks.

\subsection{Parameters Sensitivity Study (RQ4)}
\begin{figure}[htbp]
    \centering
    \subfigure[Impact of cold-start task weight]{
        \includegraphics[width=.5\linewidth]{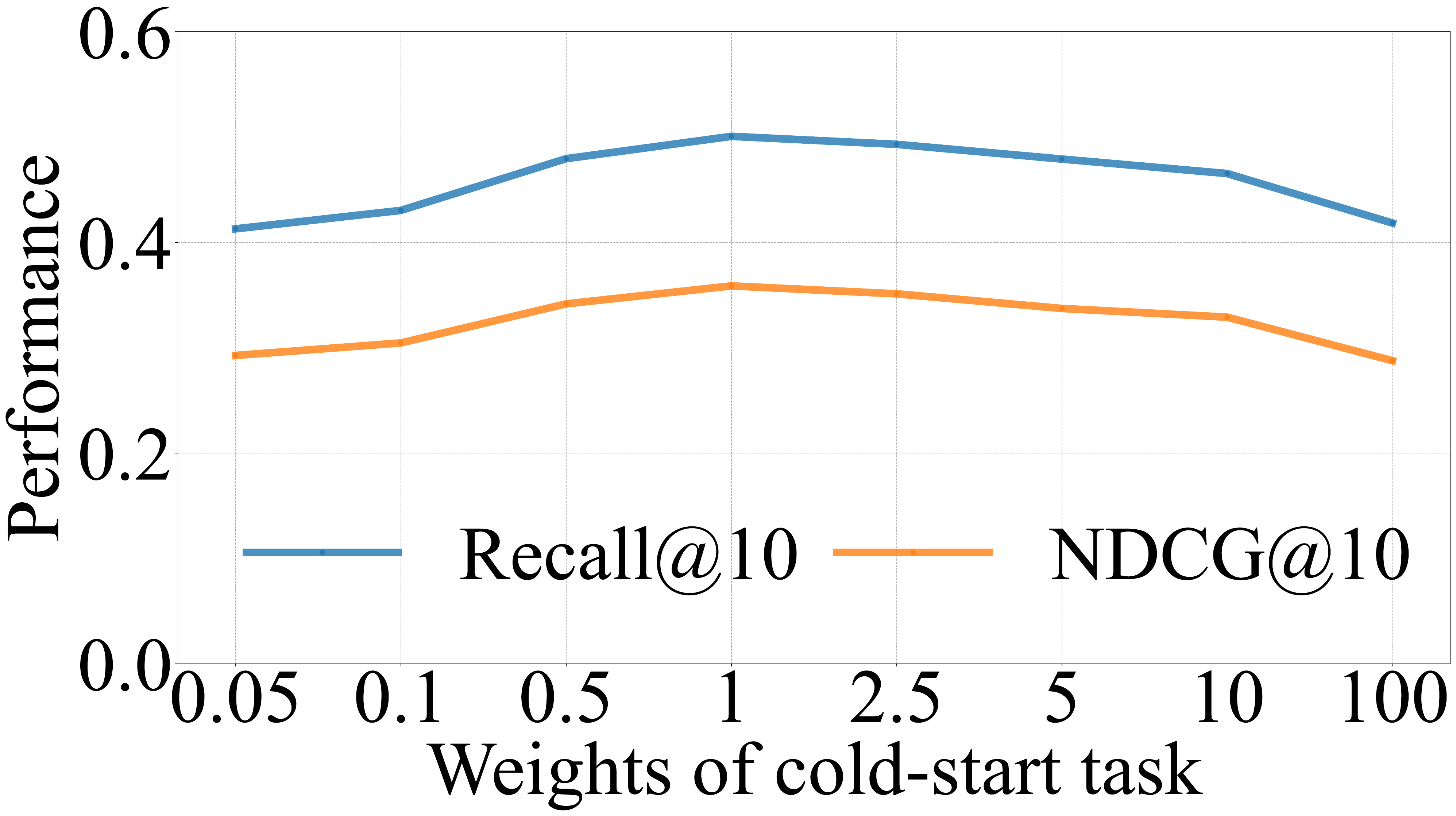}
        \label{fig:task}
    }
    \subfigure[Impact of augmentation loss weight]{
        \includegraphics[width=.47\linewidth]{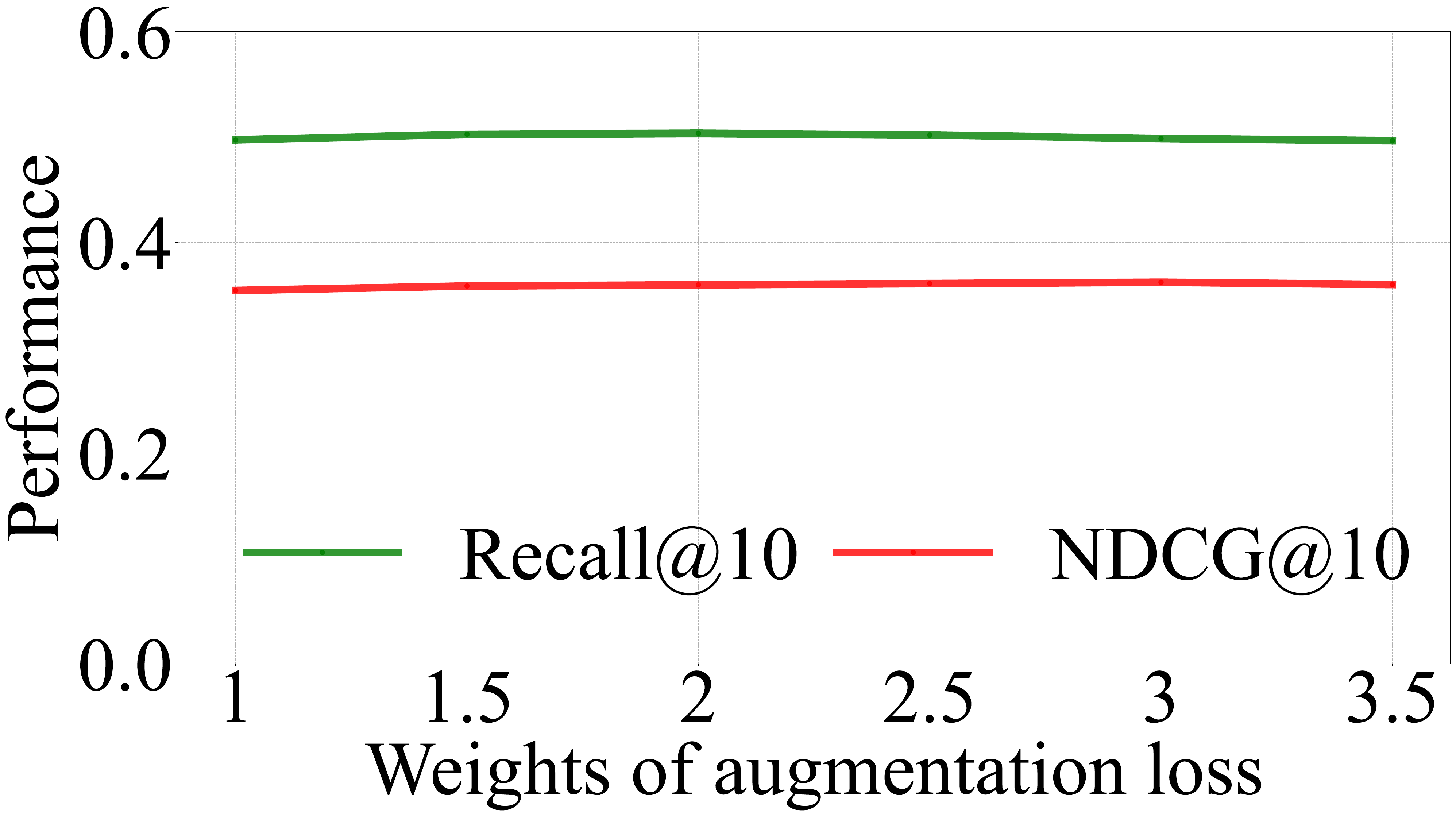}
        \label{fig:aug}
    }
    \subfigure[Impact of self-supervised loss weight]{
        \includegraphics[width=.47\linewidth]{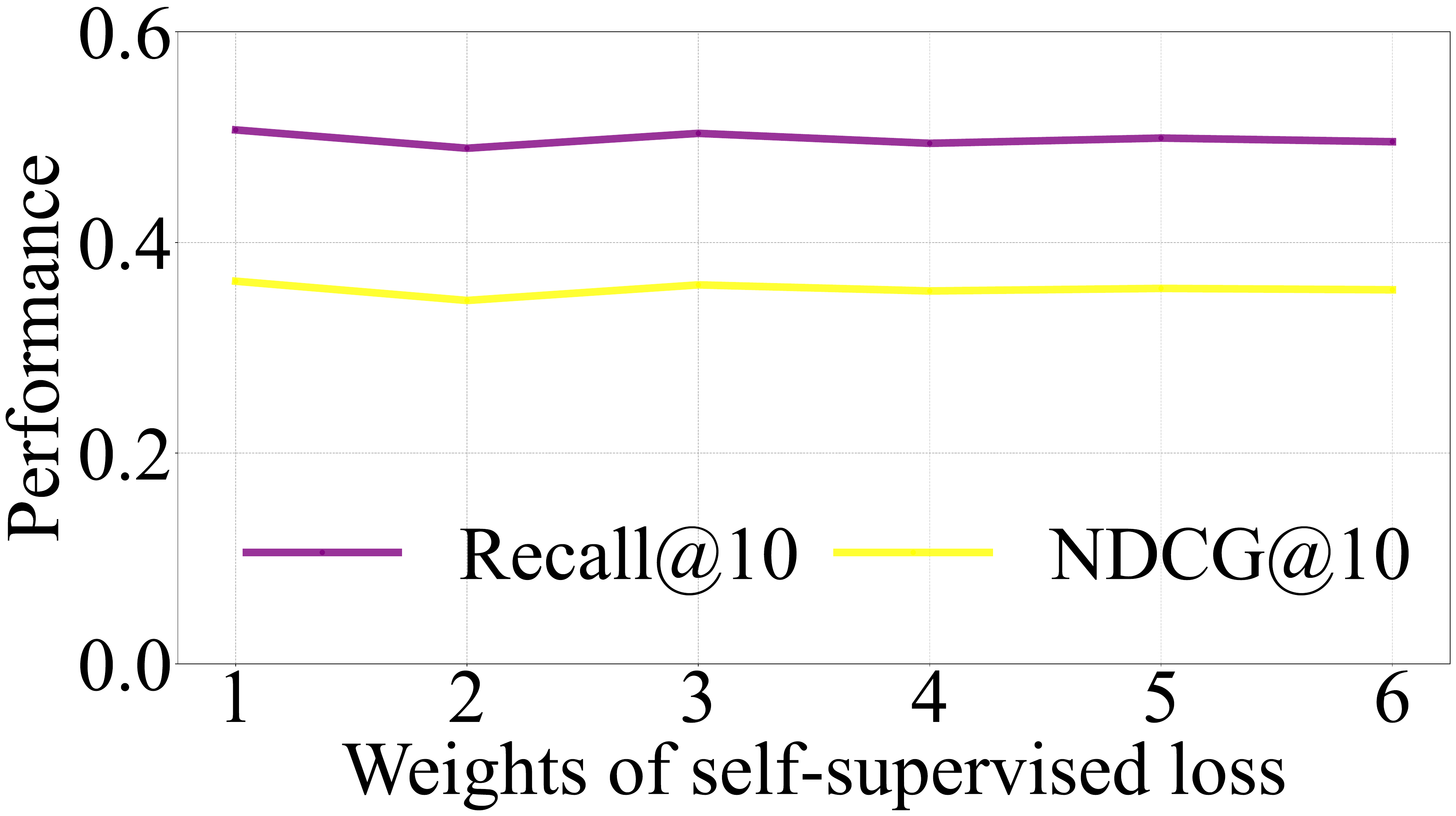}
        \label{fig:sup}
    }
    \caption{Performance of \method \textit{w.r.t.} different weights of cold-start tasks and different weights of losses.}
    \label{fig:param}
\end{figure}

We analyzed the parameter sensitivity of \method in two parameter dimensions: the weight of the cold-start task $\lambda$ and the weight between the losses of  different modules $\gamma$, the results are displayed in Fig.~\ref{fig:param}.

\subsubsection{Impact of cold-start task weight} 

We evaluated the impact of different weights for the cold-start task in calculating the loss of meta-learning in Fig.~\ref{fig:task}. It can be seen that when the weight of the cold-start task is set in an appropriate range, it can effectively improve the performance on cold-start items. If the weight of the cold-start task is set too low, the loss on the cold-start parameter will take a minor percentage of the overall loss during the global update, thus making the parameter less specialized for the cold-start task. On the other hand, if the weights of the cold-start tasks are set too high, then as mentioned before, most of the information in the popular data will be lost, leaving the parameters shared between the tasks poorly updated, leading to the same inferior performance.

\subsubsection{Impact of modules loss weight}

We similarly analyzed the effect of the loss weights of different modules on the effectiveness of \method. It can be seen that for the data-enhanced modules in Fig.~\ref{fig:aug}, the enhancement brought are more stable and the weights only have minor effect on the metric performance. Besides, the self-supervised guidance module, although performance stably, fluctuates slightly more than the data enhancement module, as shown in Fig.~\ref{fig:sup}.

\subsection{Online A/B Tests (RQ5)}
To evaluate the performance of PAM in an online system, we deployed the \method in a commercial, billion-user-scale, online recommender system and compared it to the PF baseline. 
For company privacy, we don’t report the implementation details such as the size of online data, the QPS of user requests and the real performance of the original retrieval stage. Instead, we report the performance gain ratio improved by our approach \method.
In the online system, we counted the following metrics, including the rate of items appearing in users' recommendations (Show\%), and the ratio of users liking (LTR), commenting (CMTR), and collecting videos (CLTR). The results of the experiments in Table~\ref{tab:AB} show that the high accuracy of PAM for item recommendation enables items to be better recommended to users who favor the item.

\begin{table}[t]
    \centering
    \caption{Online performance of \method compared to PF.}
    \begin{tabular}{ccccc}
        \toprule
         & Show\%& LTR& CMTR& CLTR\\\midrule
        \method & +41.39\%& +60.45\%& +4.26\%& +6.34\%\\
        \bottomrule
    \end{tabular}
    \label{tab:AB}
\end{table}

\section{Conclusion}
In this work, we find the problem of poor results of online recommender systems on cold-start items due to the long-tailed distribution of item popularity, as well as the inapplicability of existing cold-start schemes in scenarios with streaming data. To this end, we propose a unique \method by partitioning the data into tasks by item popularity and optimizing the performance of cold-start task parameters while sharing information between tasks. Furthermore, we design a novel cold-start task enhancer to optimize the performance of cold-start tasks further by leveraging the popular item information. The conducted experiments demonstrate the superiority of the \method approach, and in the future, we will further investigate different schemes to optimize the online cold-start problem.

\begin{acks}
This work is partially sponsored by National Key R\&D Program of China under Grant 2022YFB3104200, NSFC U24A20235 and 62032003. 
\end{acks}

\clearpage
\bibliographystyle{ACM-Reference-Format}
\bibliography{reference}

\appendix

\section{Datasets Settings}~\label{sec:dataset}
We selected three public datasets that containing timestamp information.

\textbf{MovieLens}~\cite{MovieLens} is a dataset of user ratings for movies. We selected one of the MovieLens 25M datasets and retained the data for the period 2014-2018 for training, containing 6,840,091 interactions between 43,181 users and 51,142 movies.

\textbf{Yelp} is a dataset containing user reviews of various businesses spanning over 10 years, containing 6,990,280 interactions between 1,987,929 users and 150,346 businesses.

\textbf{Book}~\cite{Book} is a part of the \textbf{Amazon} dataset, which collects book reviews from purchasers on the Amazon e-shopping site. We selected data from the period 2012-2023 for training and retained users with 10 or more interactions and items with 3 and more interactions, remaining 6,402,728 interactions between 351,487 users and 581,717 books.

\end{document}